\documentclass[aps,prl,twocolumn,groupedaddress,showpacs]{revtex4}
\usepackage{amssymb}
\usepackage{amsfonts}
\usepackage{amsmath}
\usepackage{graphicx}

\def\<{\langle}
\def\>{\rangle}

\def\C{{\mathbb C\, }}

\def\SG{{\mathfrak S}}

\def\i{{\rm i}}
\def\qb#1{\left|#1\right>}

\begin{document}

\title{Algebraic invariants of five qubits}

\author{Jean-Gabriel Luque}
\email{luque@univ-mlv.fr}
\author{Jean-Yves Thibon}
\email{jyt@univ-mlv.fr}
\affiliation{Institut Gaspard Monge, Universit\'e de Marne-la-Vall\'ee\\
F-77454 Marne-la-Vall\'ee cedex, France}

\date{February 15, 2005}

\begin{abstract}
The Hilbert series of the algebra of polynomial invariants of five
qubits pure states is obtained, and the simplest invariants are
computed.
\end{abstract}


\maketitle

\section{\label{intro}Introduction}
Quantifying entanglement in multipartite systems is a fundamental
issue in Quantum Information Theory.
However, for systems with more than two parts, very
little is know in this respect. A few useful entanglement measures
for pure states of $3$ or $4$ qubits have been investigated \cite{Bre, MW, Ema}, but
one is still far from a complete understanding. Furthermore,
for system of up to
$4$ qubits, a complete classification of entanglement
patterns and of corresponding invariants under local filtering
operations (also known as SLOCC, Stochastic Local Operations
assisted by Classical Communication) is know \cite{Ver,LT3}.
Klyachko \cite{Kly,KS}
proposed to associate entanglement (of pure states) in a
$k$-partite system (or perhaps , one should say ``pure
$k$-partite'' entanglement)  with the mathematical notion of
semi-stability, borrowed from geometric invariant theory,
which means that
at least one SLOCC invariant is non zero. For such states,
the absolute values of
these invariants provide some kind of entanglement measure.
However, even for
system of $k$ qubits, the complexity of these invariants grows very
rapidly with the number of parts. For $k=2$, they are given by
simple linear algebra \cite{Per,HHH}. The case $k=3$ is already nontrivial but
appears in the physics literature in \cite{DVC} and
boils down to a mathematical result which was known by 1880 \cite{LeP}.
The case $k=4$ is quite recent \cite{LT3}, and to the best of our knowledge,
nothing was known for $5$-qubit systems\footnote{Just after we posted the
first version of this Note, A. Osterloh and J. Siewert informed us
of their independent work \cite{Ost1} on the five qubits problem (see Conclusion
for a short discussion). }.

Our main result is a
closed expression of the Hilbert series of the algebra of SLOCC
invariants of pure $5$-qubit states. This result, which determines
the number of linearly independent homogeneous invariants in any degree,
was obtained through
intensive symbolic computations relying on a very recent algorithm for
multivariate residue calculations. We point out a few properties which
can be read off from the series,
and determine the
simplest invariants, which are of degree $4$ and $6$  in the
component of the states.

\section{\label{Hilbert} Hilbert series}

Denote by $V=\C^2$ the local Hilbert space of a two state particle.
The state space of a five particule  system is
${\cal H}=V^{\otimes 5}$, which will be regarded as the natural representation of the
group of invertible local filtering operations, also known as
reversible stochastic local quantum operations assisted by classical communication
$$G=G_{\rm SLOCC}={\rm SL}(2,\C)^{\times 5},$$
that is, the group of $5$-tuples of complex unimodular $2\times 2$ matrices.
We will denote by
\[
|\Psi\rangle=\sum_{i_1,i_2,i_3,i_4,i_5=0}^1A_{i_1i_2i_3i_4i_5}|i_1\rangle
|i_2\rangle|i_3\rangle|i_4\rangle
|i_5\rangle
\]
a state of the system. An element ${\bf g}=({}^kg_i^j)$ of $G$ maps
$|\Psi\rangle$ to the state
\[
|\Psi'\rangle={\bf g}|\Psi\rangle
\]
whose components are given by
\begin{equation}
A'_{i_1i_2i_3i_4i_5}=\sum_{\bf j}{}^1g_{i_1}^{j_1}
{}^2g_{i_2}^{j_2}{}^3g_{i_3}^{j_3}{}^4g_{i_4}^{j_4}{}^5g_{i_5}^{j_5}
A_{j_1j_2j_3j_4j_5}
\end{equation}

We are interested in the dimension of the
space ${\cal I}_d$ of all $G$-invariant homogeneous polynomials
of degree $d=2m$ (${\cal I}_d=0$ for odd $d$) in the 32 variables
$A_{i_1i_2i_3i_4i_5}$.

It is known that it is equal to the multiplicity of the trivial
character of the symmetric group $\SG_{2m}$ in the fifth power of
its irreducible character labeled by the partition $[m,m]$
\begin{equation}
\dim {\cal I}_d=\langle \chi^{2m}|(\chi^{mm})^5\rangle.
\end{equation}

The generating function of these numbers
\begin{equation}
h(t)=\sum_{d\ge 0}\dim{\cal I}_d\,t^d
\end{equation}
is called
the Hilbert series of the algebra ${\cal I}=\bigoplus_d {\cal I}_d$.
Standard manipulations with symmetric functions allow to express it
as a multidimensional residue:
\begin{equation}
h(t)=\oint\frac{du_1}{2\pi\i u_1}\cdots\oint\frac{du_5}{2\pi\i u_5}
\frac{A({\bf u})}{B({\bf u};t)}
\end{equation}
where the contours are small circles around the origin,
\begin{equation}
A({\bf u})=\prod_{i=1}^5\left(1+1/u_i^2\right)
\end{equation}
and
\begin{equation}
B({\bf u};t)=\prod_{a_i=\pm 1}
(1-t\,u_1^{a_1}u_2^{a_2}u_3^{a_3}u_4^{a_4}u_5^{a_5})
\end{equation}
Such multidimensional residues are notoriously difficult to evaluate.
After trying various approaches, we eventually succeded by means of a
recent algorithm due to Guoce Xin \cite{Xin}, in a {\tt Maple} implementation.
The result can be cast in the form

\begin{equation}
h(t)={P(t)\over Q(t)}
\end{equation}
where
$P(t)$ is an even polynomial of degree $104$ with non negative integer coefficients
$a_n$
\begin{equation}
P(t)=\sum_{k=0}^{52}a_{2k}t^{2k}\nonumber
\end{equation}
given in table \ref{tab1},
\begin{table}[t]
\begin{equation}
\begin{array}{|c|c|}
\hline n&a_n\\
\hline 0&1\\
8&16\\
10&9\\
12&82\\
14&145\\
16&383\\
18&770\\
20&1659\\
22&3024\\
24&5604\\
26&9664\\
28&15594\\\hline
\end{array}\,\,\
\begin{array}{|c|c|}
\hline n&a_n\\
\hline 30&24659\\
32&36611\\
34&52409\\
36&71847\\
38&95014\\
40&119947\\
42&14849\\
44&172742\\
46&195358\\
48&214238\\
50&225699\\
52&229752\\
\hline
\end{array}\,\ \,
\begin{array}{|c|c|}
\hline n&a_n\\
\hline 54&225699\\
56&214238\\
58&195358\\
60&172742\\
62&146849\\
64&119947\\
66&95014\\
68&71847\\
70&52409\\
72&36611\\
74&24659\\
76&15594\\
\hline
\end{array}\,\ \,
\begin{array}{|c|c|}
\hline n&a_n\\
\hline 78&9664\\
80&5604\\
82&3024\\
84&1659\\
86&770\\
88&383\\
90&145\\
92&82\\
94&9\\
96&16\\
104&1\\ &\\\hline
\end{array}\nonumber
\end{equation}
\caption{Coefficients of $P(t)$ \label{tab1}}
\end{table}
and $$Q(t)=(1-t^4)^5(1-t^6)(1-t^8)^5(1-t^{10})(1-t^{12})^5.$$

On this expression, it is clear that a complete description of the algebra
of $G$-invariant polynomials
by generators and relations is out of reach of any computer system.

\noindent Nevertheless, inspection of the Hilbert series suggests
the following kind of structure for this  algebra.
We know, since $\dim {\cal H}-\dim G=2^5-3\times 5=17$, that
there must exist a set of 17 algebraically independent invariants.
The denominator of the series, which is precisely a product
of 17 factors, makes it plausible that these invariants can be
choosen as five polynomials of degree 4 (to be denoted by
$D_x D_y, D_z, D_t, D_u$), one polynomial of degree $6$ ($F$),
five polynomials of degree $8$ ($H_1, H_2\dots, H_5$),
one polynomial of degree $10$ ($J$) and five polynomials
of degree $12$ ($L_1,\dots,L_5)$. These 17 polynomials
are called the primary invariants.

The numerator should then describe the secondary invariants,
that is, a set of 3014400 homogeneous polynomials
($1$ of degree $0$ , 16 of degree $8$, 9 of degree $10$,
$82$ of degree $12$ etc)
such that
any invariant polynomial can be uniquely expressed as a linear
combination of secondary invariants, the coefficients being
themselves polynomials in the primary invariants.

This picture, which is the simplest kind of description
 to be expected, is far too complex for physical applications.
The best that can be done is to use the Hilbert series as a guide
for finding explicitly a small set of reasonably simple invariants,
in particular, the primary invariants of lowest degrees.
We have computed the first primary invariants, those of degree
4 and 6, using methods from Classical Invariant Theory
(see below).

\section{The simplest invariants}

\subsection{Transvectants and Cayley's Omega process}

In order to apply the formalism of Classical
Invariant Theory, a state $|\Psi\rangle$ will
be interpreted as a quintilinear form on $\C^2$
(called the ground form)
$$f:=\sum_{i_1,i_2,i_3,i_4
i_5=0}^1A_{i_1i_2i_3i_4i_5}x_{i_1}y_{i_2}z_{i_3}t_{i_4}u_{i_5}$$
A covariant
of $f$ is a $G$-invariant polynomial in the coefficients
$A_{i_1i_2i_3i_4i_5}$ and the variables $x_i, y_i, z_i, t_i$ and
$u_i$. A complete set of covariants can be in principle computed
from the ground form by means of the so-called Omega process (see \cite{Pea,
Olv}
for notations). Cayley's Omega process consists in applying
iteratively differential operators
called transvections and defined by
\[
\begin{array}{rcl}(P,Q)^{\epsilon_1\dots\epsilon_5}&=&
{\rm
tr\ }\Omega_x^{\epsilon_1}\cdots\Omega_u^{\epsilon_5}P(x',\dots,u')
Q(x'',\dots,u'')\end{array}\]
where
\[\Omega_x=\det\left|\begin{array}{cc}\partial\over\partial {x'}_1&
\partial\over\partial {x'}_2\\\partial\over\partial {x''}_1&
\partial\over\partial {x''}_2\end{array}\right|\]
and ${\rm tr}: x',x''\rightarrow x$.

\subsection{Degree 4}
Regarding $x$ as a parameter, write
 $f$ as a quadrilinear binary form in the variables $y_i,
z_i, t_i$ and $u_i$
\[ f=\sum A^x_{i_1i_2i_3i_4} y_{i_1}z_{i_2}t_{i_3}u_{i_4}\]
It is known that such a quadrilinear form
admits an invariant of degree $2$
(called Cayley's hyperdeterminant \cite{Cay,LT1,LT2})
which is a quadratic binary form $b_x$ in the variables $x=(x_1,x_2)$.
Hence, taking the discriminant of $b_x$ one obtains
an invariant $D_x$ of degree $4$. We repeat this operation
for the other binary variables and we obtain four
other invariants $D_y$, $D_z$, $D_t$ and $D_u$.
Evaluating the appropriate Jacobians with a computer
algebra system gives the algebraic independance
of the five invariants.

\subsection{Degree 6}
We obtain the primary invariant
 of degree $6$ by a succession of transvections.
First, we compute a triquadratic covariant of degree $2$
 $$B_{22020}=(f,f)^{00101}.$$
This covariant allows to construct a cubico-quadrilinear covariant of degree $3$
 $$C_{31111}=(B_{22020},f)^{01010}$$
which gives a triquadratic polynomial of degree $4$
 $$D_{22200}=(C_{31111},f)^{10011}.$$
Hence, one obtains a quintilinear covariant of degree $5$
 $$E_{11111}=(D_{22200},f)^{11100}.$$
Finally, we find the invariant of degree $6$
 $$F=(E_{11111},f)^{11111}.$$
By computing the Jacobian, one finds that $F$ is algebraically
independent of $D_x, \dots, D_u$.

\section{\label{concl}Conclusion}
From the Hilbert series, it appears that the algebra of polynomial
invariants of a five qubit system has a very high complexity.
Furthermore, as is already the case with  smaller systems
\cite{Ver,LT2,LT3}, the knowledge of the invariants is not
sufficient to classify entanglement patterns. In the case of four
qubits or three qutrits, this classification can be achived due to
hidden symmetries which have their roots in very subtle aspects of
the theory of semi-simple Lie algebras (Vinberg's theory
\cite{VP}). However, such symmetries are absent in the case of 5
qubits. Then, the only known general approach for classifying
orbits (entanglement patterns) requires the computation of the
algebra of covariants, which is already almost intractable in the
case of four qubits. It has 170 generators, which have been found
\cite{LT3}, but the description of their algebraic relations
(syzygies) is definitely out of reach. However, a closer look at
the $4$-qubit system, reveals that the classification of
Verstraete et al \cite{Ver,Ver2}. can be reproduced by means of
only a  small set of covariants.
\begin{table}[t]
$\begin{array}{c|c|c|c|c}
&\qb{\Phi_1}&\qb{\Phi_2}&\qb{\Phi_3}&\qb{\Phi_4}\\
 \hline D_x&\times&\times&0&0\\
 \hline D_y&\times&\times&0&0\\
 \hline D_z&\times&0&0&0\\
 \hline D_t&\times&0&0&0\\
 \hline D_u&\times&0&0&0\\
 \hline F&0&0&0&0\\
 \hline B_x&\times&\times&\times&\times\\
 \hline C_{31111}&0&0&\times&\times\\
 \hline E_{11111}&0&\times&0&\times
\end{array}$
\caption{\label{tab2} Evaluation of SLOCC covariants for Osterloh
and Siewert states ($\times$ means that the evaluation is not $0$)
}
\end{table}
We hope that our results will allow the identification and the
calculation of such a small set of invariants and covariants,
sufficient to separate the physically relevant entanglement
patterns, which are probably not so numerous. To illustrate this
principle, let us consider a result of  Osterloh and Siewert
\cite{Ost1}. 
Having introduced a notion of {\it filter} which
can be used to separate SLOCC orbits in the same way as covariants,
these authors show that the four states
\begin{eqnarray}
 \qb{\Phi_1}={1\over\sqrt2}\left(\qb{11111}+\qb{00000}\right)\nonumber\\
 \qb{\Phi_2}={1\over2}\left(\qb{11111}+\qb{11100}+\qb{00010}+\qb{00001}\right)\nonumber\\
 \qb{\Phi_3}={1\over\sqrt6}\left(\sqrt2\qb{11111}+\qb{11000}+\qb{00100}+\qb{00010}\right.\nonumber\\
 \left.+\qb{00001}\right)\nonumber\\
 \qb{\Phi_4}={1\over2\sqrt2}\left(\sqrt3\qb{11111}+\qb{10000}+\qb{01000}+\qb{00100}\right.\nonumber\\
\left. +\qb{00010}+\qb{00001}\right)\nonumber
\end{eqnarray}
are in different orbits. As  can be seen on Table \ref{tab2},
the orbits of these states are also distinguished by our
covariants.

Finally, the investigation of entanglement measures
requires an understanding of
invariants under local unitary transformations (LUT)
\cite{Gra} . In a forthcoming  paper, we will explain how to
obtain LUT-invariants from SLOCC-covariants.



\bibliographystyle{unsrt}
\bibliography{5qubits5}

\end{document}